\documentclass[prl,unsortedaddress,superscriptaddress]{revtex4}
\usepackage{amsmath}  %amstex obsolete
\usepackage{amsbsy,amssymb,amsfonts}
\usepackage{graphicx}
\usepackage{color}
\usepackage{epsf}
\usepackage[utf8]{inputenc}
\usepackage{hyperref}
\usepackage{multirow}
\def\cm{$\rm cm^{-1}$}

\def\bravert{\egroup\,\vrule\,\bgroup}

{\catcode`\|=\active
  \gdef\Twoint#1{\left(\mathcode`\|"8000\let|\bravert {#1}\right)}}
{\catcode`\|=\active
  \gdef\Braket#1{\left<\mathcode`\|"8000\let|\bravert {#1}\right>}}

\newcommand{\beq}{\begin{equation}}
\newcommand{\eeq}{\end{equation}}
\newcommand{\beqa}{\begin{eqnarray}}
\newcommand{\eeqa}{\end{eqnarray}}
\newcommand{\bea}{\begin{array}}
\newcommand{\eea}{\end{array}}

\newcommand{\bef}{\begin{figure}}
\newcommand{\ef}{\end{figure}}
\newcommand{\bc}{\begin{center}}
\newcommand{\ec}{\end{center}}
\newcommand{\bt}{\begin{table}}
\newcommand{\et}{\end{table}}
\newcommand{\btb}{\begin{tabular}}
\newcommand{\etb}{\end{tabular}}

\def\molcas{$\cal M\kern-0.10em O\kern-0.15em L\kern-0.00em 
             C\kern-0.10em A\kern-0.05em S$}
\begin{document}

\vspace{2cm}
\title {{ 
         Comment on ``Theoretical study of thorium monoxide for the electron electric dipole moment search:
         Electronic properties of ${H}^3\Delta_1$ in {T}h{O}''
       }}

\vspace*{2cm}

\author{Malika Denis}
\email{malika.denis@irsamc.ups-tlse.fr}
\affiliation{Laboratoire de Chimie et Physique Quantiques,
             IRSAMC, Universit{\'e} Paul Sabatier Toulouse III,
             118 Route de Narbonne, 
             F-31062 Toulouse, France}

\author{Timo Fleig}
\email{timo.fleig@irsamc.ups-tlse.fr}
\affiliation{Laboratoire de Chimie et Physique Quantiques,
             IRSAMC, Universit{\'e} Paul Sabatier Toulouse III,
             118 Route de Narbonne, 
             F-31062 Toulouse, France}

\date{\today}
\vspace*{1cm}
%%%%%%%%%%%%%%%%%%%%%%%%%
\begin{abstract}
We present an updated EDM effective electric field of
$E_{\text{eff}} = 75.2\left[\frac{\rm GV}{\rm cm}\right]$
and the electron-nucleon scalar-pseudoscalar interaction constant $W_S=107.8$ [kHz] for the ${^3\Delta}_1$ 
science state of ThO. 
The criticisms made in reference [J. Chem. Phys. {\bf{142}}, 024301 (2015)] are addressed and largely
found to be unsubstantiated within the framework of our approach.
\end{abstract}

\maketitle

\section{Introduction}
%%%%%%%%%%%%%%%%%%%%%%%%%%%%%%%%%%%%%%%%%%%%%%%%
\label{SEC:INTRO}

Recent experimental \cite{ACME_ThO_eEDM_science2014} and theoretical \cite{Fleig2014,Skripnikov_ThO_JCP2013} 
studies on the ThO molecule have led to a new and improved upper bound on the electron electric dipole moment
(eEDM), $d_e$. This upper bound is determined through $d_e = -\frac{\hbar\omega^{\cal{NE}}}{E_{\text{eff}}}$,
where $\omega^{\cal{NE}}$ is an upper bound to a measured frequency shift and $E_{\text{eff}}$ is the
EDM effective electric field, {\it{i.e.}}, it is the combined result of a measurement and a molecular
many-body calculation. Since the theoretical uncertainty for $E_{\text{eff}}$ enters the upper bound on
$d_e$ directly, this uncertainty should be minimized.

However, most accurate results for the required EDM effective electric field $E_{\text{eff}}$ in the 
${^3\Delta}_1$ science state of ThO from two different approaches (Skripnikov {\it{et al.}} \cite{Skripnikov_ThO_JCP2015} 
and Fleig {\it{et al.}} \cite{Fleig2014}) are at variance by $6.3 \left[\frac{\rm GV}{\rm cm}\right]$, or about
8\%. Furthermore, it is alleged by Skripnikov {\it{et al.}} \cite{Skripnikov_ThO_JCP2015} that the error bars given
in reference \cite{Fleig2014} were significantly underestimated. 

In the present comment, we address the criticism advanced by Skripnikov {\it{et al.}} in reference \cite{Skripnikov_ThO_JCP2015}
through an additional elaborate study, we present an improved value of $E_{\text{eff}}$ for ThO (${^3\Delta}_1$)
and the value of the electron-nucleon scalar-pseudoscalar (enSPS) interaction constant.
The latter is determined as described in reference \cite{ThF+_NJP_2015} and represents the second leading ${\cal{P,T}}$-odd 
effect in ThO, allowing to constrain the electron-nucleon coupling $C_S$.

\section{Results and Discussion}
%%%%%%%%%%%%%%%%%%%%%%%%%%%%%%%%%%%%%%%%%%%%%%%%
\label{SEC:APPL}

\subsection{Spinor basis set}
There are physically reasonable and physically unreasonable choices for the spinor basis in a correlation
model which falls short of Full CI. Skripnikov {\it{et al.}} \cite{Skripnikov_ThO_JCP2015} include a
less reasonable choice in their determination of the sensitivity of the MR-CI method with respect to spinor
basis, namely ground-state (${^1\Sigma^+}$) spinors. Naturally, the inclusion of such unmotivated choices
will lead to arbitrarily large error bars, in the extreme case. For instance, any random excited state
could also have been chosen for determining the spinor basis.

Instead, we only use physically well motivated choices for spinor basis which in the present case are the 
following: i) DCHF spinors with an average-of-occupation Fock operator for 2 electrons in 3 Kramers pairs,
$7s_{\sigma}$ and $6d_{\delta}$, model DCHF\_$2$in$3$. Such a basis is not state specific but gives a 
balanced description of the ground ${^1\Sigma^+}$ and the excited ${^{1,3}\Delta}$ states which is an 
advantage in the determination of energetics. ii) DCHF spinors with 1 electron occupying $7s_{\sigma}$ 
and 1 electron occupying $6d_{\delta}$, model DCHF\_$1$in$1$\_$1$in$2$. This latter model is specific
towards the excited ${^{1,3}\Delta}$ states and better suited for a property calculation in the ${^3\Delta_1}$
science state.

The comparative results are compiled in Table \ref{TAB:SPINORS}. Not surprisingly, the excitation energy
of $\Omega=1$ depends strongly on whether the $\delta$ spinors are included in the DCHF averaging or not.
The hyperfine interaction constant also undergoes changes of a few percent.
However, $\cal{P,T}$-odd properties are almost totally insensitive to the choice of spinors. Our final
results from reference \cite{Fleig2014} were based on DCHF\_$2$in$3$ spinors which are at variance from the
state-specific spinors by not more than $0.3$\%. Furthermore, $\cal{P,T}$-odd constants are also insensitive
to basis set enlargement within the 4c-MR$_{12}$-CISD(18) model, in contrast to what has been asserted by
Skripnikov and Titov.
\begin{table}[h]
\caption{\label{TAB:SPINORS}
         Calculated properties for $\Omega=1$ at $R = 3.477$ a$_0$, using the wavefunction model MR$_{12}$-CISD(18), the
         vDZ basis set and a virtual cutoff value of 50 a.u. The results using the same correlation model and the vTZ 
         basis set (see reference \cite{Fleig2014}) have been added for comparison.
        }

\begin{center}
\begin{tabular}{l|rrrr}
 Spinor basis         & $T_v$ [\cm] & $E_{\text{eff}} \left[\frac{\rm GV}{\rm cm}\right]$ & $A_{||}$ [MHz] & $W_S$ [kHz] \\ \hline
 DCHF\_$2$in$3$ (vTZ)     & $5410$  &   $75.2$                                            & $-1339$         & $105.8$         \\
 DCHF\_$2$in$3$           & $6069$  &   $75.1$                                            & $-1333$         & $105.3$         \\
 DCHF\_$1$in$1$\_$1$in$2$ & $6066$  &   $74.9$                                            & $-1291$         & $105.1$         \\ 
 DCHF\_cs                 & $7871$  &   $75.0$                                            & $-1375$         & $105.4$
\end{tabular}
\end{center}
\end{table}

\subsection{Active spinor space}

We have carried out an additional study to confirm sufficient convergence of our results with respect to the
size of the active spinor space. Results are compiled in Table \ref{TAB:ACTIVE}. To this end, we have further
increased the parameter $K$ given in Fig. 1 of reference \cite{Fleig2014} to values which group types of
spinors in accord with their principal atomic character. The active space corresponding to $K=31$ includes
spinors up to an energy of $0.527$ a.u.  
\begin{table}[h]

\caption{\label{TAB:ACTIVE}
         Calculated properties for $\Omega=1$ at $R = 3.477$ a$_0$, using different active spinor spaces ($X$) with the wavefunction
         model MR$_{X}$-CISD(18) and vDZ basis set with a virtual cutoff of 50 a.u.
        }

\begin{center}
\begin{tabular}{r|rrr}
Model                & $E_{\text{eff}} \left[\frac{\rm GV}{\rm cm}\right]$ & $A_{||}$ [MHz] & $W_S$ [kHz] \\ \hline
MR$_{3}$-CISD(18)    &              $80.8$         &  $-1283$             &   $113.7$       \\
vTZ/MR$_{3}$-CISD(18)    &          $81.0$         &  $-1292$             &   $114.1$       \\    
%     MR$_{5}$-CISD(18)    &          $78.9$                               &                 \\
%     MR$_{6}$-CISD(18)    &          $77.0$                               &                 \\
%     MR$_{8}$-CISD(18)    &          $75.4$                               &                 \\
MR$_{9}$-CISD(18)    &              $73.8$         &  $-1321$             &   $103.7$       \\
MR$_{12}$-CISD(18)   &              $74.7$         &  $-1341$             &   $105.0$        \\
vTZ/MR$_{12}$-CISD(18)   &          $75.2$         &  $-1339$             &   $106.0$       \\
MR$_{13}$-CISD(18)   &              $74.7$         &  $-1343$             &   $104.9$       \\
vTZ/MR$_{13}$-CISD(18)   &          $75.2$         &  $-1343$             &   $105.9$       \\
MR$_{17}$-CISD(18)   &              $74.8$         &  $-1334$             &   $105.2$       \\
MR$_{31}$-CISD(18)   &              $73.1$         &  $-1320$             &   $102.7$        \\
\end{tabular}
\end{center}
\end{table}
First, we note that the characteristic drop of $E_{\text{eff}}$ (and also $W_S$) occurs largely independent
of basis set extent, which is in accord with the analysis of this effect presented in reference 
\cite{Fleig2014}. Upon increasing the active space to $K=31$, we observe a further slight decrease of the
$\cal{P,T}$-odd constants. The corresponding configuration space adds a large number of triple and quadruple
excitations to spaces with smaller value of $K$. These quadruples are of the type 
$occ^{16}val^{2} \longrightarrow occ^{14}{val*}^{2}virt^{2}$ where the superindex is an occupation number, 
the occupied space ($occ$) comprises the
Th 6s,6p and the O 2s,2p shells, the valence space is divided into Th 7s,6d$\delta$ ($val$) and spinors
below an energy of $0.527$ a.u. ($val*$), and the virtual space ($virt$) represents all spinors of higher
energy.

\subsection{Core correlations}
The correction to the $\cal{P,T}$-odd properties by the inclusion of even more inner-shell electrons in the 
correlation treatment was studied in a previous work \cite{Fleig2014} and estimated to be $0.25\%$ by the 
comparison of the MR$_{3}$-CISD(18) and MR$_{3}$-CISD(36)* models. However, it was pointed out in reference
\cite{Skripnikov_ThO_JCP2015} that the 36-electron calculation was performed with a smaller cut-off of virtual 
spinors of 5 Hartrees. Hence, by determining within their 2c-CCSD(T) framework that the truncation leads to a 
$3.3$ GV/cm underestimation, Skripnikov {\it{et al.}} asserted that the uncertainty due to the number of explicitly 
correlated electrons amounts to $5\%$.

\begin{figure}[h]
\caption{\label{FIG:VIRTUAL}
Calculated properties (arbitrary units) for $\Omega=1$ at $R = 3.477$ a$_0$, using the wavefunction
         model MR$_{3}$-CISD(36), vTZ basis set and different cutoff values for the virtual spinor
	 space.
	 }

\vspace{0.5cm}

 \begin{center}
  \includegraphics[width=14.0cm,angle=0]{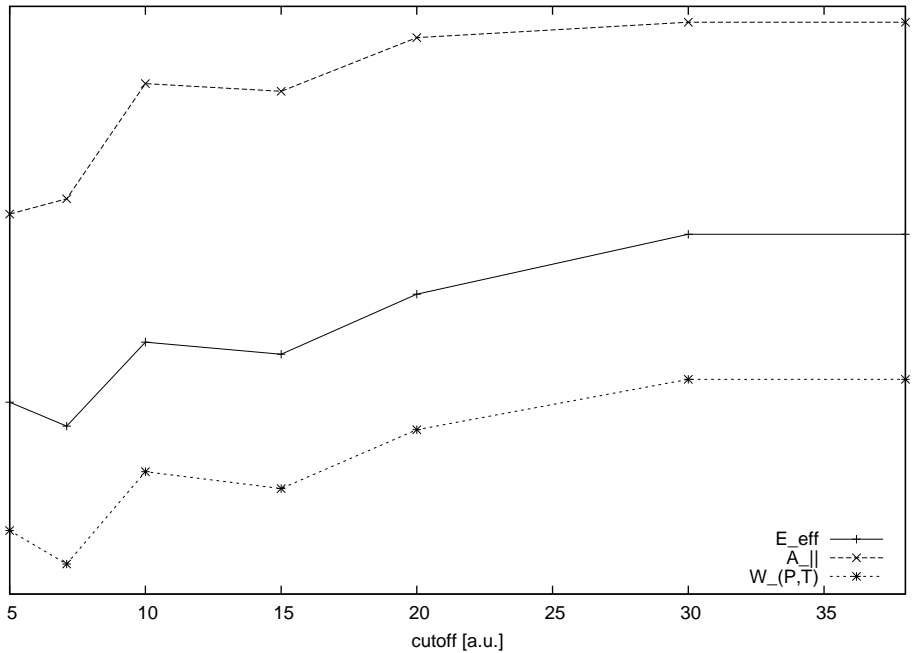}
 \end{center}
\end{figure}

In order to check this figure, we carried out a study of the effect of the truncation of the virtual space
for the MR$_{3}$-CISD(36) model. Results are shown graphically in Figure \ref{FIG:VIRTUAL}.
It appears that indeed convergence is not reached at the 5 Hartree cut-off level but values are accurate when 
we apply a 30 Hartree truncation. Hence, in \cite{Fleig2014} the values of the $\cal{P,T}$-odd properties in 
the MR$_{3}$-CISD(36)* model were underestimated by $1.7\%$ at the most.
The expansion of the virtual spinor space is accompanied by an increase of $E_{\text{eff}}$, $W_S$ and $A_{||}$ 
on the absolute. 
The effect is strongest when adding p-type spinors to the virtual space.
%and consequently a greater s-p mixing
%In our case, it results in an enhancement of the $\cal{P,T}$-odd interactions.
Besides, this study led us to perform the calculation of the properties for a 38 Hartrees cut-off, the same as 
for the 18-electron model in \cite{Fleig2014}. Therefore, the correction on the effective electric field, coming 
from the correlation of more core electrons can be determined with accuracy. It amounts to $+1.2$ GV/cm, {\it{i.e.}},
an increase of $1.5\%$ in magnitude, which is significantly smaller than the $+4.3$ GV/cm alleged by Skripnikov 
{\it{et al.}} \cite{Skripnikov_ThO_JCP2015}.

\subsection{Subvalence and valence correlations}
In order to start from more rigorous base values, the calculation of the $\cal{P,T}$-odd properties was performed within the vTZ/MR$_{12}^{+T}$-CISD(18) model 
that corresponds to the MR$_{K}^{+T}$-CISD(18) model defined in \cite{Fleig2014} with an active space of $12$ Kramers pairs
and the use of vTZ basis sets. This model differs from the previous reference model vTZ/MR$_{12}$-CISD(18) by 
allowing for three holes in the Th 6s, 6p and O 2s 2p subvalence spinors.  
%the inclusion of triple excitations from the Th 6s, 6p and O 2s 2p subvalence spinors ($occ$) to the active space. 
In particular, this model includes a subset of Quintuple excitations deriving from the following types of excited configurations: 
$occ^{16}val^{2} \longrightarrow occ^{13}{val*}^{3}virt^{2}$, $occ^{16}val^{2} \longrightarrow occ^{13}{val*}^{4}virt^{1}$, 
$occ^{16}val^{2} \longrightarrow occ^{13}{val*}^{5}virt^{0}$ where the $occ$, $val$ and $virt$ spaces are the same as defined above
and $val*$ comprises the Th 7p, 8s, 8p$_\pi$ spinors. 
The inclusion of these higher excitations from the  subvalence spinors ($occ$) to the active space
entails an increase  of the values of $2.5\%$ in magnitude for the $\cal{P,T}$-odd properties, leading to the new base values to which will be added the various corrections discussed above.

\subsection{Gaunt operator}
Finally, so as to account for the Gaunt interaction, the Gaunt term was added to the Dirac-Coulomb Hamiltonian. 
This was possible at the Hartree-Fock level for which the Dirac-Coulomb-Gaunt Hamiltonian is implemented \cite{sikkema_MMF} in the
\verb+DIRAC+ program. Thus, the only $\cal{P,T}$-odd property implemented at this level, {\it{i.e.}}, 
the EDM effective electric field ($E_{\text{eff}}$), 
was evaluated as an expectation value of the operator over the Hartree-Fock spinors. 
Details on the implementation of the EDM operator can be found in reference \cite{fleig_nayak_eEDM2013}.
For the evaluation, we employed the same state-specific model DCHF\_$1$in$1$\_$1$in$2$ as decribed above
that is the most adequate for the calculation of the properties in the ${^3\Delta_1}$ molecular term. 
The comparison of the values of $E_{\text{eff}}$ without and with the inclusion of the Gaunt operator shows 
a non-negligible decrease of $1.7\%$ in magnitude.
 
\section{Conclusion}
%%%%%%%%%%%%%%%%%%%%%%%%%%%%%%%%%%%%%%%%%%%%%%%%
\label{SEC:SUMM}

In this work, we tackled criticisms made by Skripnikov {\it{et al.}} \cite{Skripnikov_ThO_JCP2013}. 
The main point was the alleged underestimated uncertainty on E$_{\text{eff}}$ due to the choice of the 
spinor basis ($7$\%); 
yet, our work revealed the insensitivity of $\cal{P,T}$-odd properties to proper choices of spinor space.
Second, based on the analysis of their 18-electron MR($\infty$)-CISD model, Skripnikov et al. asserted that our previous final value obtained by an MR(12)-CISD 
calculation could undergo a significant increase of 5$\%$ in magnitude. Thus, even if the non-relativistic MR($\infty$)-CISD and our four-component
MR(12)-CISD cannot be compared straightforwardly, we addressed this particular issue through two studies. 
A review of the effect of the size of the active space led to a correction of -1.6
$\left[\frac{\rm GV}{\rm cm}\right]$. Second,
in order to refine our understanding of the subvalence and valence correlations, we included higher excitations 
through the MR$_{12}^{+T}$-CISD(18) model.
We came to perform a 7-billion determinant CI calculation that yielded new reference values. 
%Despite the lack of explicit quadruple excitations, 
The latter model includes a subset of quadruple and even quintuple excitations with respect to the ground-state
reference determinant
besides the triple excitations from the subvalence to the active space. 
Furthermore, the influence of the inclusion of core electrons in the correlation space was accurately 
analyzed by correlating up to 36 electrons 
and resulted in an increase of $E_{\text{eff}}$ by $+1.2 \left[\frac{\rm GV}{\rm cm}\right]$.
A survey of the Gaunt interaction brought about an additional correction of 
$-1.3 \left[\frac{\rm GV}{\rm cm}\right]$. 
All corrections are compiled in Table \ref{TAB:THO:PROPERCORR}. 
\begin{table}[h]

\caption{\label{TAB:THO:PROPERCORR}
         Final property values including corrections
        }

\vspace*{0.5cm}
\hspace*{-1.7cm}
\begin{center}
\begin{tabular}{rrr|l}
$E_{\text{eff}} \left[\frac{\rm GV}{\rm cm}\right]$ & $A_{||}$ [MHz]  & $W_S$ [kHz] \\ \hline
 $75.2$\footnotemark[1] &  $-1339$  &    $106.0$ &   vTZ/MR$_{12}$-CISD($18$)  \\ \hline
$77.1$             &  $-1309$  &  $108.5$ &   new base value from vTZ/MR$_{12}^{+T}$-CISD($18$)  \\
$-0.2$                  &  $+42$    &  $-0.2$  &   correction for $\Delta$ spinors \\
$-1.6$                  &  $+21$    &  $-2.3$  &   correction for active space size \\
$+1.2$                  &  $-20$    &  $+1.8$  &   core correlations \\
$-1.3$                  &  &    &   Gaunt correction  \\ \hline
$75.2$                  &  $-1296$ &  $107.8$ & {\bf{Final value}} \\ \hline
\end{tabular}
\footnotetext[1]{Reference \cite{Fleig2014}}
\end{center}
\end{table}
Based on this study, we propose improved values of the EDM effective electric field 
$E_{\text{eff}} = 75.2\left[\frac{\rm GV}{\rm cm}\right]$
and the electron-nucleon scalar-pseudoscalar interaction constant $W_S=107.8$ [kHz] for the ${^3\Delta}_1$ 
science state of ThO. 
%\malika{We may conclude on the uncertainty, how do you determine it ?}
The corrections we have deduced are within our previously assigned uncertainty of $3$\%. Furthermore, our
present final value of $E_{\text{eff}} = 75.2\left[\frac{\rm GV}{\rm cm}\right]$ is within the uncertainty
margins of the combined results from references \cite{Fleig2014} and \cite{Skripnikov_ThO_JCP2013}.

\begin{acknowledgments}
%%%%%%%%%%%%%%%%%%%%%%%%%%%%%%%%%%%%%%%%%%%%%%%%
 %\input{acknowledgments}

We thank Trond Saue (Toulouse) for a valuable suggestion and Malaya K. Nayak (Mumbai)
for helpful discussions.
Financial support from the {\it{Agence Nationale de la Recherche}} (ANR) through grant no.
ANR-BS04-13-0010-01, project ``EDMeDM'', is gratefully acknowledged.
\end{acknowledgments}

%%%%%%%%%%%%%%%%%%%%%%%%%%%%%%%%%%%%%%%%%%%%%
%                                           %
%         Appendices                        %
%                                           %
%%%%%%%%%%%%%%%%%%%%%%%%%%%%%%%%%%%%%%%%%%%%%
%\newpage
%\begin{appendix}
% \input{appendix}
%\end{appendix}

%\clearpage

%%%%%%%%%%%%%%%%%%%%%%%%%%%%%%%%%%%%%%%%%%%%%
%                                           %
%         Bibliography                      %
%                                           %
%%%%%%%%%%%%%%%%%%%%%%%%%%%%%%%%%%%%%%%%%%%%%
\bibliographystyle{unsrt}
%\bibliography{main}
%\bibliography{all}
%\input main.bbl

\newcommand{\Aa}[0]{Aa}

\end{document}